\documentclass[a4paper,10pt]{article}
\usepackage{graphicx}
\usepackage{mathptmx}      
\usepackage{amssymb}
\usepackage{amsmath}
\usepackage{authblk}
\usepackage{color,colortbl}

\definecolor{lightRed}{RGB}{230,230,255}
\newcommand{\vp}{\varphi}

\newcommand{\pd}{\partial}

\begin{document}

\title{Voltammetry: mathematical modelling and Inverse Problem}

\author{N.A.Koshev}
\affil{
		Institute of Computational Mathematics, 
		University of S\~{a}o Paulo, 
		S\~{a}o Carlos, SP 13566-590, Brazil, 
              	\emph{nikolay.koshev@gmail.com}           
}

\author{A.N.Koshev}
\author{V.V.Kuzina}

\affil{
		Penza State University of Architecture and construction, 
		Penza, 440028, Russian Federation,
		\emph{ankoshev@ramler.ru}
}

\maketitle

\begin{abstract}
	We propose the fast semi-analytical method of modelling the polarization curves 
	in the voltammetric experiment. The method is based on usage of the special functions and 
	shows a big calculation speed and a high accuracy and stability. 
	Low computational needs of the proposed algorithm allow us to state the set of 
	Inverse Problems of voltammetry for the reconstruction of metal ions concentrations or 
	the other parameters of the electrolyte under investigation. 
\end{abstract}

\section{Introduction}\label{sec:intro}
The methods of the voltammetric measurements and the mathematical processing of them are often used 
for a wide set of the electro-chemical and physical problems, such as: the definition of metal ions 
in electrolytes, refinement or definition of the other parameters of the processes in 
electrochemical systems. Such methods are based on usage of the polarization phenomenon for 
obtaining and interpreting the \emph{polarization curves} - the dependencies of the system current on 
the applied voltage. Variety of the systems for voltammetric research leads to a big number of the 
mathematical methods of a processing the obtained information (see \cite{GALUS}-\cite{BOOK_VOLT}). The most popular and 
effective of them are based on the regression analysis or on the research of physico-mathematical models 
of the diffusion processes and the processes near the electrodes. 

In the experimental sense, the regression analysis is based on usage of the automatic periodic update of the electrode working surface 
and programmable controllers of generating the potential sweeps (\cite{REGR}). Polarization of the electrode with the linearly changing 
potential, registration of the voltammetric dependencies and statistical processing supposed to be done 
automatically (\cite{REGR}). The object function (\emph{regression dependence}) refers to the dependence of the currents and the potentials 
of the peaks of the voltammetric curves on the metal ion concentrations. Such dependencies are obtaining using the 
factor experiment planing.  
The process of building such dependencies is a very labour intensive and complicated. For each private case 
we need to make a huge number of experiments, which is necessary for the statistical building of the 
regression models. Furthermore, there is a threat of loosing the accuracy on the three steps: experiment, 
data processing and solving of the Inverse Problems of definition of concentrations or other parameters. 

On the other side, the most simple methods of the determined modelling of voltammetric curves are 
often based on the private cases. For example, in \cite{LSM} the analytical description of the 
polarization curves is based on the least square method. 
For very slow potential sweeps one can also solve quasi-stationary diffusion equation \cite{SPRINGER-2015}. 
In general case, we have to solve the system of equations, describing diffusion and electrode processes. However, 
the solution of these equations with the classical methods is rather complicated and needs 
big computational resource, especially when the big number of model is needed. 

We propose fast semi-analytical method of modelling the voltammetric curves 
with usage of special functions. The possibility of fast numerical modelling allows us to state
the Inverse Problem of voltammetry, main purpose of which is the definition of voltammetry 
process parameters (including metal ion concentrations). 

The diffusion and electrode processes in voltammetry can be described with the following system of
parabolic type equations (PDEs). 
\begin{equation}\label{initial_eq}
	\begin{aligned}
		& \frac{\pd C^k}{\pd t} = D_k\frac{{\pd}^2 C^k}{\pd x^2}\text{,} \\
		& C^k(x, 0) = C_0^k, \quad C^k(\delta_k, t) = C_0^k, \quad 
			\frac{\pd C^k}{\pd x}(0, t) = \frac{i^k(t)}{Z_k FD_k}, \\
		& i(t) = \sum\limits_{k=1}^{K}i^k(t) \quad  t\in[0, T],\quad x\in[0, \delta], \quad k = 0,1,...,K-1, \\
	\end{aligned}
\end{equation}
where index $k = 1,2,...,K-1$ denotes the number of electro-active component of water solution of 
electrolytes, $C^k(x, t)$ - mass-transfer of $k$ component, $i^k(t)$ - partial current, 
$i(t)$ - shared current of the system (polarization curve), $D_k, Z_k$ - electrochemical 
constants (values, corresponding to different substances can be found in electrochemical tables, for example in 
\cite{SPRAV}),  
$\delta_k$ - thickness of the diffusion layer of $k$ component of the system, $C_0^k$ - the concentration
of $k$ component outside the diffusion layer and $T$ - time of research of the process. 

Due to \cite{GALUS}, partial currents $i_k$ and system potential $E(t)$ are related with the 
following formula: 
\begin{equation}\label{current_1}
	i^{k}(t) = i^{k0}\Big( \frac{C^k(0, t)}{C_0^k}e^{\frac{\alpha_k 
		Z_k F}{RT}(E(t) - E_0^k)} - e^{\frac{(\alpha_k-1) Z_k F}
		{RT}(E(t) - E_0^k)} \Big),
\end{equation}
where $E_0^k$ - threshold potential of $k$ system component and $i^{k0}, \alpha_k$ 
- current of exchange and transfer coefficient, which correspond to $k$ system component.

\emph{Note} that during the experiment we obtain the dependence $i(E)$, which is the definition of polarization curve. 
However, potential sweep $E(t)$ is a linear function of time $t$. For convenience, without any loss of 
generality we are changing the polarization curve $i(E)$ with $i(t)$. 

For brevity, we introduce the following notations:

\begin{equation}\label{short_not1}
	N^k(t) = e^{\frac{\alpha Z_k F}{RT}(E(t) - E_0^k)}, \quad
	R^k(t) = e^{\frac{(\alpha-1) Z_k F}{RT}(E(t) - E_0^k)}.
\end{equation}
Due to these notations, (\ref{current_1}) will take a form:
\begin{equation}\label{current}
	i^{k}(t) = i^{k0}\Big( \frac{C^k(0, t)}{C_0^k} N(t) - R(t) \Big),
\end{equation}

From these equations we can allocate two problems: direct and inverse. The first one is 
the modelling of the polarization curve with all known parameters. The second (Inverse Problem) 
is a problem of definition of some parameters on base of known polarization curve and system 
\ref{initial_eq}.

\section{Finite-dimensional approximation of voltammetric curve}\label{sec:fda_methods}
\subsection{General approach}
Let the experimental curve $i(t)$ is given on some uniform mesh $t_j \in [0, T], 0 \leq j \leq N - 1$.
Since the current of the system, obtained experimentally is a sum of partial 
currents of corresponding elements of the system, we represent it in the following form:
\begin{equation}\label{i_sum}
	i_j \equiv i(t_j) = \sum\limits_{k=0}^{K-1} i_j^k, 
\end{equation}
where $i_j^k \equiv i^k(t_j)$ - values of partial currents in the given nodes $t_j$. 

\textbf{Statement.} \emph{Let the function $i^k(t)$ can be presented on interval $t\in[0, T]$  
as a linear combination }
\begin{equation}\label{i_approx}
	i^k(t) = \sum\limits_{m=0}^{N-1} a_m l_m(t),
\end{equation}
\emph{where $a_m \in R^1$ and $l_m(t), m=0,1,...,N-1$ - some continuous functions.}

\emph{Then the solution $C^k(x, t)$ of the equations (\ref{initial_eq}) can be written in form:}
\begin{equation}\label{C_approx}
	C(x,t) = C_0 + \sum\limits_{m=0}^{N-1} a_m \vp_m^k(x,t), 
\end{equation}
\emph{ where functions $\vp(x,t)$ are the solutions of the following problem:}

\begin{equation}\label{vp_eq}
	\begin{aligned}
		& \frac{\pd \vp_m^k}{\pd t} = D\frac{\pd^2 \vp_m^k}{\pd x^2}; 
		\quad \vp_m^k(x, 0)=0; \quad \vp_m^k(\delta, t) = 0; \\
		& \frac{\pd \vp_m^k}{\pd x}(0, t) = \frac{l_m(t)}{Z_k F D_k}, \quad t\in[0,T]
	\end{aligned}
\end{equation}

We note that the proof of this statement can be easily done with usage the uniqueness of the solution 
of the diffusion problem; one can see this proof in \cite{SPRINGER-2015}. 

For brevity we omit the indexes $k$ in further consideration. 

\begin{equation}\label{vp_eq1}
	\begin{aligned}
		& \frac{\pd \vp_m}{\pd t} = D\frac{\pd^2 \vp_m}{\pd x^2}; 
		\quad \vp_m(x, 0)=0; \quad \vp_m(\delta, t) = 0; \\
		& \frac{\pd \vp_m}{\pd x}(0, t) = \frac{l_m(t)}{Z F D}, \quad t\in[0,T]
	\end{aligned}
\end{equation}
 
This problem was considered and solved in \cite{BUDAK}. General solution can be written in form 

\begin{equation}\label{phi2}
	\begin{aligned}
		\vp(x,t) = -\frac{1}{ZF}\int\limits_{0}^{t}l_m(\tau)G(x, 0, t-\tau)d\tau, 
	\end{aligned}
\end{equation}
where function $G(x, \xi, t)$ is a source function: 
$$
	G(x, \xi, t) = \frac{1}{\delta} + \frac{2}{\delta} \sum\limits_{n=1}^{+\infty} 
		e^{-\frac{n^2 \pi^2 D}{\delta^2}t} cos\frac{n\pi x}{\delta} sim\frac{n \pi \xi}{\delta}.
$$
Thus, due to the fact that we are interested only in $\vp(0,t)$, the solution of (\ref{vp_eq1}) 
can be written in form:
\begin{equation}\label{PHI_sol}
	\begin{aligned}
		& \vp(0,t) = -\frac{1}{\delta ZF}\int\limits_{0}^{t}l_m(\tau) 
		\Big( 1 + 2\sum\limits_{n=1}^{+\infty} e^{-\gamma_n (t-\tau)} \Big) d\tau = \\ 
		& -\frac{1}{\delta ZF} \Big( I_m(t) + 2\sum\limits_{n=1}^{+\infty} J_m^n(t) \Big),
	\end{aligned}
\end{equation}
where
\begin{equation}\label{IJ_PHI}
	I_m(t) = \int\limits_{0}^{t} l_m(\tau) d\tau, \quad 
		B_m^n(t) = \int\limits_{0}^{t} l_m(\tau) e^{-\gamma_n (t-\tau)} d\tau, \quad
		\gamma_n = \frac{n^2 \pi^2 D}{\delta^2}. 
\end{equation}

\subsection{Piecewise-linear approximation}\label{sec:pw_approx}

Consider the following functions: 
\begin{equation}\label{l_odd}
	l_m^{even}(t) = \left\{ 
		\begin{aligned} 
			& \frac{t_{m+1} - t}{t_{m+1} - t_m}, \text{if} \quad t\in[t_m, t_{m+1}] \\
			& 0, \text{if} \quad t\notin [t_m, t_{m+1}]
		\end{aligned}
		\right.
\end{equation}
\begin{equation}\label{l_even}
	l_m^{odd}(t) = \left\{ 
		\begin{aligned} 
			& \frac{t - t_{m-1}}{t_m - t_{m-1}}, \text{if}\quad t\in[t_{m-1}, t_m] \\
			& 0, \text{if}\quad t\notin [t_{m-1}, t_m]
		\end{aligned}
		\right.
\end{equation}

On each interval $t\in[t_m, t_{m+1}]$ curve is presented with the sum of increasing and decreasing 
linear functions: 
\begin{equation}
	i(t) = i_m l_m^{even}(t) + i_{m+1} l_{m+1}^{odd}(t),	
\end{equation}
and, on the whole interval of interest  $t\in[0, T]$: 
\begin{equation}\label{i_lin_approx}
	\begin{aligned}
		& i(t) = i_0 l_0^{even}(t) + i_1 l_1^{odd}(t) + i_1 l_1^{even}(t) + i_2 l_2^{odd}(t) + ... +
			i_{N-1}^{odd}(t) = \\
		& = i_0 l_0^{even}(t) + \sum\limits_{m=1}^{N-2} i_m 
			\big( l_m^{odd}(t) + l_m^{even}(t) \big) + i_{N-1} l_{N-1}^{odd}(t) 
			= \sum\limits_{m=0}^{N-1} i_m l_m(t),
	\end{aligned}
\end{equation}
where 
\begin{equation}\label{I_m}
	l_m(t) = \left\{ 
		\begin{aligned} 
			& l_0^{odd}(t), & \text{if}\quad m=0, \\
			& l_m^{odd}(t) + l_{m}^{even}(t), & \text{if}\quad m=1,2,...,N-2, \\
			& l_{N-1}^{odd}(t), & \text{if}\quad m=N-1. \\
		\end{aligned}
		\right.
\end{equation}

Using the expressions \ref{I_m}, we can write $B_m^n$ from \ref{IJ_PHI}: 
\begin{equation}
	B_m^n(t) = \left\{ 
		\begin{aligned} 
			& B_0^{n(even)}(t), & \text{if}\quad m=0, \\
			& B_m^{n(odd)}(t) + B_{m}^{n(even)}(t), & \text{if}\quad m=1,2,...,N-2, \\
			& B_{N-1}^{n(odd)}(t), & \text{if}\quad m=N-1. \\
		\end{aligned}
		\right.
\end{equation}
where $B_m^{n(odd)}$ and $B_m^{n(even)}$ are the integrals of functions $l_m^{odd}$ and $l_m^{even}$. 

Taking into account \ref{l_odd} and \ref{l_even} we can write:
$$
B_m^{n(odd)}(t_l) = \left\{
	\begin{aligned}
		& 0, & l\le m-1 \\ 
		& \frac{1}{d}\int\limits_{t_{m-1}}^{t_m}(\tau - t_{m-1}) e^{-\gamma_n(t_m - \tau)}d\tau, & l \ge m 
	\end{aligned} 
	\right.
$$

$$
B_m^{n(even)}(t_l) = \left\{
	\begin{aligned}
		& 0, & l\le m \\
		& \frac{1}{d}\int\limits_{t_{m}}^{t_{m+1}}(t_{m+1} - \tau) e^{-\gamma_n(t_{m+1} - \tau)}d\tau, & l \ge m+1
	\end{aligned}
	\right.
$$
We introduce the notations:
\begin{equation}\label{U_PHI}
	\begin{aligned} 
		U_m^n \equiv U^n = \int\limits_{t_{m-1}}^{t_m} e^{-\gamma_n(t_m - \tau)} d\tau = 
		\frac{1 - e^{-\gamma_n d}}{\gamma_n}
	\end{aligned}
\end{equation}

\begin{equation}\label{V_PHI}
	\begin{aligned} 
		V_m^n = \int\limits_{t_{m-1}}^{t_m} \tau e^{-\gamma_n t_m - \tau} d\tau 
		= \frac{1}{\gamma_n} (t_m - t_{m-1}e^{-\gamma_n d} - U^n).	
	\end{aligned}
\end{equation}
thus, $B_m(t)$ can be calculated using the following expressions: 
\begin{eqnarray}\label{B_PHI}
	B_m^n(t_l) = \left\{ 
		\begin{aligned}
			& 0, & l < m, \\
			& \frac{1}{d}(V_m^n - t_{m-1} U^n), & l = m; \\
			& \frac{1}{d}(V_m^n - V_{m+1}^n + 2d U^n), & l>m;
		\end{aligned}
		\right. \nonumber \\ 
	B_0^n(t_l) = \left\{ 
		\begin{aligned}
			& 0, & l = 0, \\
			& \frac{1}{d}(t_1 U^n - V_1^n), & l > 0; 
		\end{aligned}
		\right. \\ 
	B_{N-1}^n(t_l) = \left\{ 
		\begin{aligned}
			& 0, & l \le N-2, \\
			& \frac{1}{d}(V_{N-1}^n - t_{N-2} U^n), & l=N-1. 
		\end{aligned}
		\right. \nonumber
\end{eqnarray}
In this case it is easy to calculate the integral $I_m(t)$. Functions $l_m^{odd}(t)$ and $l_m^{even}(t)$ 
on intervals $t\in[t_m, t_{m+1}]$ are linear and non-negative and the integral can be calculated as a square of
the area under the curves. Thus, 

\begin{eqnarray}\label{I_exact}
	I_0(t_l) = \left\{ 
		\begin{aligned}
			& 0, & l = 0, \\
			& \frac{d}{2}, & l > 0; 
		\end{aligned}
		\right. \quad 
	I_m(t_l) = \left\{ 
		\begin{aligned}
			& 0, & l < m, \\
			& \frac{d}{2}, & l = m; \\
			& d, & l>m;
		\end{aligned}
		\right. \nonumber \\ 
	I_{N-1}(t_l) = \left\{ 
		\begin{aligned}
			& 0, & l \le N-2, \\
			& \frac{d}{2}, & l=N-1. 
		\end{aligned}
		\right. 
\end{eqnarray}

Substituting (\ref{B_PHI}) and (\ref{I_exact}) to (\ref{PHI_sol}) we can obtain functions $\vp_m(0,t)$.
This method is fast and undemanding to the computational resource because of existence of the analytical 
formulae for integral $J_m^n(t)$. Thus, it is advisable to use this method in iterative calculation of 
the parameters under investigation, which needs repeated solving of the direct problem of modelling 
voltammetric curves (this method will be considered below). Method of piecewise-linear approximation of 
the voltammetric curves is suitable for most of modelling and Inverse Problems of voltammetry. As some 
disadvantages we can highlight less accuracy of the approximation of strongly-nonlinear parts of the 
voltammetric curves, but this is rather rare type of voltammetric curves. Also, we note, that this 
disadvantage is easily being coped with some oversampling of measured data. 

\section{The Inverse Problem}

In this section we assume the polarization curve $i(t)$ to be known - obtained with 
voltammetric experiment. The aim of Inverse Problem is to define some of parameters
$C_0^k, \alpha_k, i_0^k, D_k, E_0^k$ using the curve and equations (\ref{initial_eq}-\ref{current}). 

\subsection{General approach.} Let us define the formal vector $\Omega$, which consists 
of the parameters to be found. For example, if we need to find parameters $C_0^k$ and $\alpha_k$, 
vector $\Omega = \{ C_0^0, C_0^1,...,C_0^{K-1}, \alpha_0, \alpha_1,...,\alpha_{K-1} \}$. 
We define the \emph{cost functional} $S(\Omega)$ as a difference between modelled and 
experimental curves:

\begin{equation}\label{cost_func}
	S(\Omega) = ||i^{\Omega}(t) - i(t)||_{B}^2, 	
\end{equation}
where $i^{\Omega}(t)$ - the current, modelled using parameters from vector $\Omega$, 
$B$ - some Banach space. The space $B$ can be designed to converge algorithm faster, or 
to do some regularization (for example, in future we plan to build (\ref{cost_func}) as a 
strongly convex functional, see \cite{CARLEMAN}), but in this work we use the space $B=L_2(0,T)$.
We call \emph{approximate solution} a vector $\Omega^*$, 
on which functional (\ref{cost_func}) reaches its minimum, i.e. the parameters, minimizing the 
difference between modeled and experimental curve. Thus, the Inverse Problem is reduced to 
the optimization problem. 
Of course, to solve this problem we need firstly prove an existence, uniqueness 
and stability of its solution. In general case such proof is a complicated mathematical problem. 
However, mostly we need only find the concentrations $C_0^k$, currents $i_0^k$ and transmitting 
coefficients $\alpha_k$. For this private cases existence and uniqueness of such problem can be easily 
proved (see some proofs in \cite{KOKUZ}). Further we consider only these cases. Despite the fact that we do not present 
here strict mathematical research of stability of approximate solution, our numerical experiments show, 
that it is stable. We propose two approaches to the optimization problem for the functional (\ref{cost_func}). 

\subsection{Analytical optimization.} The cost functional (\ref{cost_func}) is a quadratic functional, which 
allows us to find its minimum analytically using the equation $\nabla S(\Omega) = 0$, where $\nabla$ denotes a 
gradient. Such approach can be very fast and elegant. However, this approach is not flexible, because we 
have to find analytical solution 
for each configuration of formal vector $\Omega$. Below we show 
very briefly the analytical solution for $\Omega = \{ C_0 \}$. This solution considered in details in \cite{KOKUZ} 
and presented here just to explain the main idea of the analytical approach.

\subsubsection{Analytical approach for single-component task.}
Let us consider the single-component problem. For brevity, we omit component indexes $k$. 
We construct the cost functional using the variable $\xi = 1/C_0$: 
\begin{equation}
	S_{inf}(\xi) = || i^{\xi}(t) - i(t) ||_{L_2}^2,
\end{equation}
where $i(t)$ is the experimental curve and $i^{\xi}(t)$ is the curve, modelled for given $\xi$. On 
finite mesh this function takes the following form:
\begin{equation} \label{func_1d}
	S(\xi) =\sum\limits_{l=0}^{N-1} \Big( i_l^{\xi} - i_l \Big)^2,
\end{equation}
Here $f_l = f(t_l)$ for any function $f(t)$. 

For brevity of further discussion, we add one more notation to the group (\ref{short_not1}):
\begin{equation}
	L(t) = \sum\limits_{m=0}^{N-1}i_m\vp_m(0,t), 
\end{equation}
and, referring to (\ref{C_approx}), rewrite the expression (\ref{current}) in the following form: 
\begin{equation}
	i(t) = i^{0}(1+\xi L(t))N(t) - R(t). 
\end{equation}

The approximate solution $\xi$ is a minimizer of (\ref{func_1d}):
\begin{equation}\label{anal_sol}
	\xi = \frac{\sum\limits_{l=0}^{N-1} L_l N_l \big( i_l+i_0(R_l - N_l) \big) }
		{i_0 \sum\limits_{l=0}^{N-1} L_l^2 N_l^2}, 	
\end{equation}
We derived this formula in \cite{KOKUZ}.

Note that if we need to find another parameters, the analytical solution will be presented in another form. 
However, the idea is the same. 

As advantages of this approach: 
\begin{itemize}
	\item Big calculation speed and low calculation resource needs - we do not need to solve the 
		equation (\ref{initial_eq}).
	\item Simplicity of the implementation. 
\end{itemize}
Disadvantages: 
\begin{itemize}
	\item Low stability to noises in experimental curve. 	
	\item Low flexibility - if we need to reconstruct other parameters, we have to obtain 
		another formulas. In case of searching on several parameters, the analytical solution
		can be complicated and cumbersome. 
\end{itemize}

\subsubsection{Analytical approach for Multi-component task.}  
Multi-component task can be easily reduced to the set of a single-component problems. Consider for simplicity  
double-component task ($K=2$) with the component threshold potentials $E_0^1 > E_0^0$. We highlight two important 
points: 
\begin{itemize}
	\item[1.] The system current $i(t)$ is a sum of the partial currents; 
	\item[2.] The partial current $i^k(t) = 0$ when $t: E(t) < E_0^k$.
\end{itemize}
These points allow us to allocate a part ($t: 0 \le E(t) \le E_0^1$) of the polarization curve, which represents only 
one electro-active component. Thus, considering this part of the curve, we can state the single-component 
problem for the component $k=0$. After solution of this problem using the formula (\ref{anal_sol}), we can build 
a model $i^0(t)$ for the first component. The curve of the single-component problem for the second component can be 
obtained with subtraction: $i^1(t) = i(t) - i^0(t)$. 

However, in real situations, the threshold potentials can be 
rarely separated enough clear for using this approach. Moreover, the disadvantages of analytical approach more 
affects the solution in case of $K>1$ because of using for each component calculation the part of the 
experimental curve, often lying in area of weak currents. Such task requires the research of ill-posedness and 
developing some regularization algorithms. In this article, despite a good potential, we do not 
describe this approach. 

\subsubsection{Iterative approach.}
The essence of this approach is searching the minimum of the cost functional using iterative 
optimization. In general, the cost functional can be built as a norm on different Banach spaces, which allows 
to apply regularization procedures (see \cite{BIBLE}), clarify the solution or stability to errors in experimental curves 
or table parameters. In this article we consider norm $L_2$, in which (\ref{cost_func}) takes the form:
\begin{equation}
	S(\Omega) = ||i^{\Omega}(t) - i(t)||_{L_2}^2, 	
\end{equation}
The advantage of this approach is a flexibility with respect to the parameters to be found. Below, in 
numerical examples, we show 
the efficiency of searching two parameters (instead of one) for the single-component problem. Iterative 
optimization allows to include in calculation as many parameters, as we need. Also, such optimization 
is flexible to iterative optimization methods, such as Hooke and Jives method (\cite{HJM}), Newton method, Conjugated 
gradients \cite{NUMOPT} and other. 

For solving multi-component problem we propose to use iterative methods. The only disadvantage of an 
iterative approach is lower calculation speed. However, this deficiency does not look very serious because of 
fast development of computing equipment. In our numerical tests we discovered that calculations spend split 
seconds using rather old laptop processor. In our numerical research we used Hooke and Jeeves method and Conjugate 
gradients; however, the best result was obtained with Hooke and Jeeves algorithm. Thus, all results, presented in
our article, were obtained with this optimization.

\section{Numerical tests}
This section shows some numerical examples of parameters reconstruction. In subsection "Model tasks" we 
consider result of the reconstruction of curves, modelled with known parameters using straight solution 
of (\ref{initial_eq}). 
All presented results were obtained using standard personal computer with processor 
Intel Core-i3.

\subsection{"Experimental" data for model tasks.}
The curves, used as an experimental data for model 
tasks, were modelled using straight solution of (\ref{initial_eq}) with all known parameters. To obtain this 
solution we used implicit Finite-Differential schemes (\cite{FDM}) on uniform mesh with size ($1000 \times 10000$). 
Such dense mesh was used to increase the stability of implicit scheme and to reduce the error in modelled curve to 
the minimum. To make situation more real, we added white noise with amplitude $10\%$ of maximum 
current value. The form of such noise corresponds to real measurements, and the amplitude was 
increased to demonstrate the stability of the method.

Since all constants are known for model task, we can estimate an accuracy of our reconstruction. 

\subsection{Single-component model task.}
In this subsection we present the results of modelling and solution of an Inverse Problem for model 
single-component task. Voltammetric curve, used an experimental, defined by Table \ref{tab:1}.

\begin{figure}[h]
	\begin{center}
		\begin{tabular}{cc}
			\includegraphics[width = 0.45\linewidth]{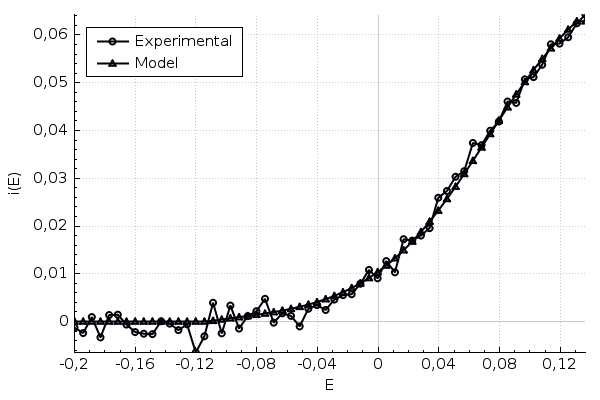} & 
			\includegraphics[width = 0.45\linewidth]{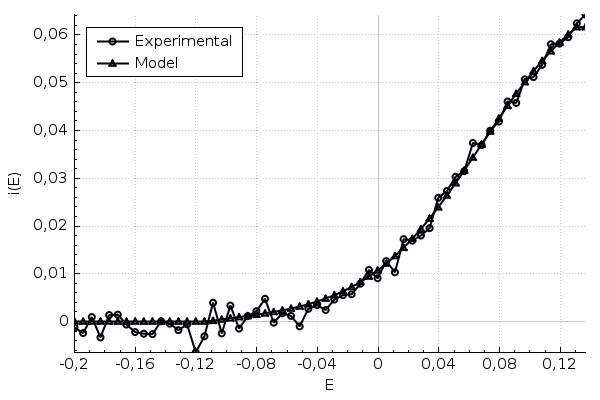} \\
			a) & b)
		\end{tabular}
	\end{center}
	\caption{\small \emph{Results of modelling and reconstruction for noised model task, defined in Table \ref{tab:1}. 
		On both pictures we use as an experimental curve the result of modelling using the 
		straight solution of (\ref{initial_eq}) with added white additional noise ($10\%$).
		a) Curve, modelled with special functions after reconstruction of parameter $C_0$ only. Calculation error: $3.44\%$. 
		b) Curve, modelled with special functions after reconstruction of parameters $C_0$ and $\alpha$. Calculation error: $1.05\%$ 
		and $1.6\%$ for parameters $C_0$ and $\alpha$ respectively. }
	\label{fig:1comp_model}
	}
\end{figure}

\begin{table}[h]
	\caption{\emph{Electrochemical parameters for single-component model \label{tab:1}}}
	\begin{center}
		\begin{tabular}{cc}
			\hline
			Parameter &\quad\quad Value \\
			\hline
			$Z$ &  \quad\quad$2$ \\
			\rowcolor{lightRed}
			$D$ &  \quad\quad $7.5e-06$ \\
			$\alpha$ &  \quad\quad$0.32$ \\
			\rowcolor{lightRed}
			$E_0$ &  \quad\quad$-0.11$ \\
			$i^0$ &  \quad\quad$0.000708$ \\
			\hline
			\hline
		\end{tabular}
	\end{center}
\end{table}
The result of solving of two Inverse Problems for single-component task is presented on Fig.\ref{fig:1comp_model}. We minimized 
of the cost functional using Hooke and Jeeves algorithm. 
During the calculations we got one interesting observation: involving in calculations two parameters (instead of one) can sufficiently 
increase the accuracy of reconstruction. More precisely, we obtained calculation error $3.44\%$ reconstructing only one parameter, and 
errors $1.05\%$ and $1.6\%$ for two parameters. Moreover, some parameters (for example, $i_0$) can be calculated only together 
with other parameters - our calculation failed while computing only $i_0$. 

\subsection{Double-component model task.}

For this test we used enter data (curve), modelled using electrochemical parameters, shown in Table \ref{tab:2}.

During this calculations we also obtained the same effect: efficiency of calculation of two 
parameters in time ($C_0$ and $\lambda$) is higher in comparison with one-parameter calculation. 
More precisely, we obtained error $19\%$ and $16.2\%$ respectively, calculating only parameters $C_0^0$ and 
$C_0^1$. 

\begin{table}[hb]
	\caption{\emph{Electrochemical parameters for double-component model \label{tab:2}}}
	\begin{center}
		\begin{tabular}{ccccc}
			\hline
			Parameter & Values for the first component & Values for the second component \\
			\hline
			$Z$ &2 &2 \\
			\rowcolor{lightRed}
			$D$ &6.1e-06 &6.5e-06 \\
			$\alpha$ &0.144 &0.32 \\
			\rowcolor{lightRed}
			$E_0$ &-0.18 &-0.05 \\
			$i^0$ &0.0064 &0.0079 \\
			\hline
			\hline
		\end{tabular}
	\end{center}
\end{table}

\begin{figure}[t]
	\begin{center}
		\begin{tabular}{cc}
			\includegraphics[width = 0.43\linewidth]{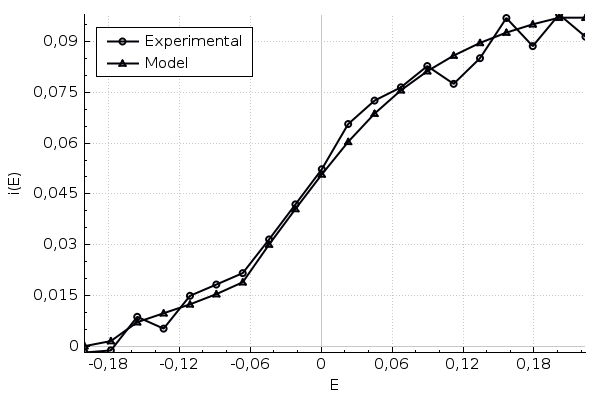} & 
			\includegraphics[width = 0.43\linewidth]{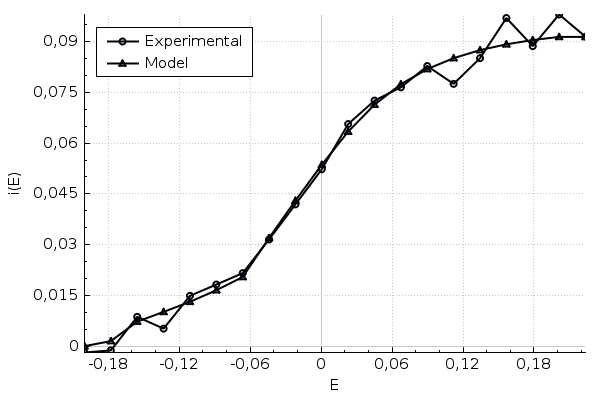} \\
			a) & b)
		\end{tabular}
	\end{center}
	\caption{\small \emph{Results of modelling and reconstruction for noised model task, defined in Table \ref{tab:2}. 
		On both pictures we use as an experimental curve the result of modelling using the 
		straight solution of (\ref{initial_eq}) with added white additional noise ($10\%$).
		a) Curve, modelled with special functions after reconstruction of parameters $C_0^k$ only. 
		Calculation errors: $19\%$ 	and $16.2\%$ for $C_0^0$ and $C_0^1$ respectively. 
		b) Curve, modelled with special functions after reconstruction of parameters $C_0$ and $\alpha$. 
		Calculation error: $3.6\%$, $6.6\%$ for 
		$C_0^0$, $C_0^1$ and $14\%$, $1.2\%$ for $\alpha^0$ ,$\alpha^1$ }
	\label{fig:2comp_model}
	}
\end{figure}

In case, when parameters $\alpha^k$ also have being calculated, we obtained errors $3.6\%$ and $6.6\%$ for
concentrations $C_0^0$ and $C_0^1$ and $14\%$ and $1.2\%$ for transmission numbers $\alpha^0$ and $\alpha^1$. 
Enter curve, modelled for this task (with parameters from Table\ref{tab:2}), and the results of modelling for 
reconstructed parameters are presented on Fig.\ref{fig:2comp_model}. 


\subsection{Single-component real task.}
\begin{figure}[ht]
	\begin{center}
		\begin{tabular}{cc}
			\includegraphics[width = 0.45\linewidth]{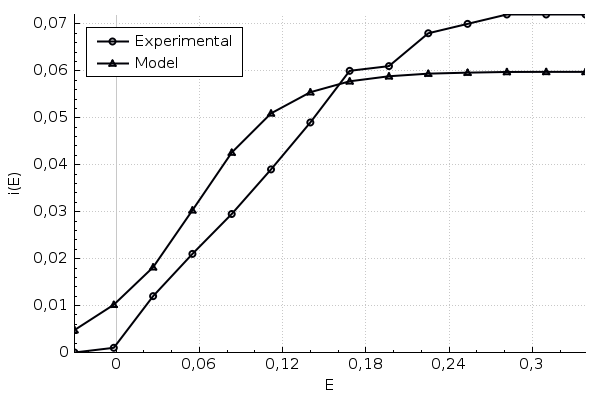} & 
			\includegraphics[width = 0.45\linewidth]{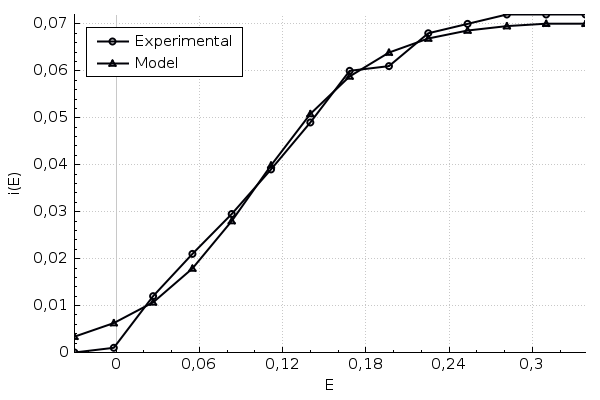} \\
			a) & b)
		\end{tabular}
	\end{center}
	\caption{\small \emph{Results of reconstruction for real task (parameters in Table \ref{tab:1}). 
		a) Curve, modelled with special functions after reconstruction of parameter $C_0$ only. 
		Calculation error: $19.2\%$.
		b) Curve, modelled with special functions after reconstruction of parameters $C_0$ and $\alpha$. 
		Calculation error: $4.6\%$, $12.2\%$ for $C_0$ and $\alpha$ respectively. } 
	\label{fig:real}
	}
\end{figure}

In this subsection we present the results of real voltammeric curve processing. The experimental data are obtained 
during the voltammetry for Cu electrolyte. All parameters are already presented in Table \ref{tab:1} (above we used 
these parameters for single-component model task). Because of knowing these parameters, we also can estimate an accuracy of 
the reconstruction. As above, the accuracy of reconstruction of two parameters is higher than the accuracy of 
one parameter reconstruction. In first case we obtained an error $19.2\%$ for reconstructed value of concentration 
$C_0$, and $4.6\%$ and $12.2\%$ for parameters $C_0$ and $\alpha$ in case of two parameters calculation. 
Experimental and modelled curves for both cases are presented on Fig.\ref{fig:real}

\bibliography{my}{}
\bibliographystyle{unsrt}
\end{document}